\documentclass[reprint,%
aps,pra,%
10pt,%
superscriptaddress,%
amsmath,amssymb,%
floatfix%
]{revtex4-2} 

\usepackage{amsmath,amssymb,bm}
\usepackage{array}
\usepackage{booktabs}
\usepackage{esvect}
\usepackage{graphicx}
\usepackage{mathtools}
\usepackage{multirow}
\usepackage{comment}
\usepackage[svgnames]{xcolor}
\usepackage[colorlinks=true,%
            linkcolor=MediumBlue,%
            urlcolor=MediumBlue,%
            citecolor=MediumBlue]{hyperref}
\usepackage{siunitx}
\renewcommand{\vec}[1]{\ensuremath{\mathbf{#1}}}

\newcommand{\fft}[1]{\ensuremath{\mathcal{F}[ #1 ]}}

\begin{document}

\title{Librational Feedback Cooling}

\author{Charles P. Blakemore}
\email{cblakemo@stanford.edu}
\affiliation{Department of Physics, Stanford University, Stanford, California 94305, USA}

\author{Denzal Martin}
\thanks{Now at: Los Alamos National Laboratory, Los Alamos, NM}
\affiliation{Department of Physics, Stanford University, Stanford, California 94305, USA}

\author{Alexander Fieguth}
\affiliation{Department of Physics, Stanford University, Stanford, California 94305, USA}

\author{\\Nadav Priel}
\affiliation{Department of Physics, Stanford University, Stanford, California 94305, USA}

\author{Gautam Venugopalan}
\affiliation{Department of Physics, Stanford University, Stanford, California 94305, USA}

\author{Akio Kawasaki}
\thanks{Now at: National Metrology Institute of Japan (NMIJ), National Institute of Advanced Industrial Science and Technology (AIST), 1-1-1 Umezono, Tsukuba, Ibaraki 305-8563, Japan}
\affiliation{Department of Physics, Stanford University, Stanford, California 94305, USA}
\affiliation{W. W. Hansen Experimental Physics Laboratory, Stanford University, Stanford, California 94305, USA\looseness=-1}

\author{Giorgio Gratta}
\affiliation{Department of Physics, Stanford University, Stanford, California 94305, USA}
\affiliation{W. W. Hansen Experimental Physics Laboratory, Stanford University, Stanford, California 94305, USA\looseness=-1}

\date{\today}

\begin{abstract}

Librational motion, whereby a rigid body undergoes angular oscillation around a preferred direction, can be observed in optically trapped, silica microspheres. We demonstrate the cooling of one librational degree of freedom for $\sim \SI{5}{\micro\meter}$ diameter spheres that have been induced to rotate with an external electric field coupled to their electric dipole moment. Cooling is accomplished by adding a phase modulation to the rotating field. The degree of cooling is quantified by applying a $\pi/2$ shift to the phase of the electric field and fitting the resulting exponential decay of the librational motion to obtain a damping time, as well as estimating a mode temperature from the observed libration in equilibrium. The result is an important step in the study of the dynamics of trapped microspheres, crucial to cooling the mechanical motion to its ground state, as well as providing insights regarding the charge mobility in the material at microscopic scales.  

\end{abstract}

\maketitle

\section{INTRODUCTION}
\label{sec:introduction}

Classical mechanical systems akin to the canonical mass-on-a-spring have been used to study oscillator dynamics under a wealth of different conditions, allowing for these systems to serve as underlying models for a variety of complex physical processes.  In the flourishing field of optomechanics~\cite{levitodynamics_review}, restoring forces are generated by optical interactions such as radiation pressure, and can thus be controlled with great precision. Indeed, translational motion of optomechanical oscillators has been cooled to the level of single quanta of the associated potential~\cite{Delic:2020, Tebbenjohanns:2020, Magrini:2021}.

To date, much less attention has been given to the rotational degrees of freedom. These have been manipulated primarily with two mechanisms: transfer of angular momentum from a circularly polarized trapping beam to a birefringent particle confined within the trap~\cite{Arita:2013, Monteiro:2018}, or coupling an external rotating electric field to the electric dipole moment within the trapped particle~\cite{Rider:2019, Blakemore:2020}. 

For the electrostatic technique, the particle must have an electric dipole moment, which has generally been observed in silica microspheres (MSs) synthesized via the St{\"o}ber process~\cite{Rider:2016, Monteiro:2018, Rider:2019, Blakemore:2020}. As the dipole moment (and thus the MS carrying it) is driven into rotation by the electric field, its orientation oscillates about the instantaneous direction of the field, which we refer to as ``libration'', in analogy with the more familiar astronomical phenomenon~\cite{Eckhardt:1981}.

We present the first demonstration of feedback cooling of a librational degree of freedom, using an optically trapped silica MS in vacuum. The MS is first translationally confined with active feedback, electrically neutralized, and then induced to rotate at a fixed, but freely chosen, angular velocity, by application of a rotating electric field.  For sufficiently small oscillations, the librational degree of freedom can be described as a damped harmonic oscillator. Feedback is provided by first detecting the phase of the MS's rotation from the polarization of transmitted light, sensitive to the rotation of the MS's birefringent axes, and subsequently modulating the phase of the rotating electric field.

The rotational motion of microscopic objects may provide systems with inherently low levels of damping~\cite{Hoang:2016, Kuhn:2017}, offer gyroscopic stabilization of the rotor's translational degrees of freedom~\cite{Arita:2013, Nagornykh:2017, Rider:2019}, as well as possibly mitigate systematic effects observed in precision force measurements with optically levitated systems~\cite{Moore:2014, Rider:2016, Blakemore:2021a, Priel:2022}.

\section{EXPERIMENTAL APPARATUS}
\label{sec:apparatus}

The optomechanical system implemented in this work consists of a vertically-oriented optical tweezers in vacuum. Silica MSs~\cite{bangs_laboratories} with diameter \SI{4.70\pm0.08}{\micro\meter}~\cite{Blakemore:2019b} are trapped at the focus of a linearly-polarized laser beam with vacuum wavelength \SI{1064}{\nano\meter} and focused with numerical aperture of ${\rm NA} = 0.12$, generated by a Yb:doped fiber laser and manipulated by a combination of both fiber and free-space optics. Translational motion of the MS in the horizontal plane is observed in the deflection of transmitted light, while vertical motion is derived from the phase of light retroreflected by the MS. Active stabilization of the translational degrees of freedom is applied by piezoelectrically-driven deflection of the trap position for the horizontal degrees of freedom, as well as power modulation of the trapping beam for the vertical degree of freedom. We observe negligible coupling between the translation and rotation of the MS, that are quite separated in frequency, $\mathcal{O}( \SIrange{10}{1000}{\hertz} )$ for translation and $\mathcal{O}( \SI{10}{\kilo\hertz} )$ for rotation. The investigation of such a coupling may prove interesting for future work.

\begin{figure}[t]
    \centering
    \includegraphics[width=1\columnwidth]{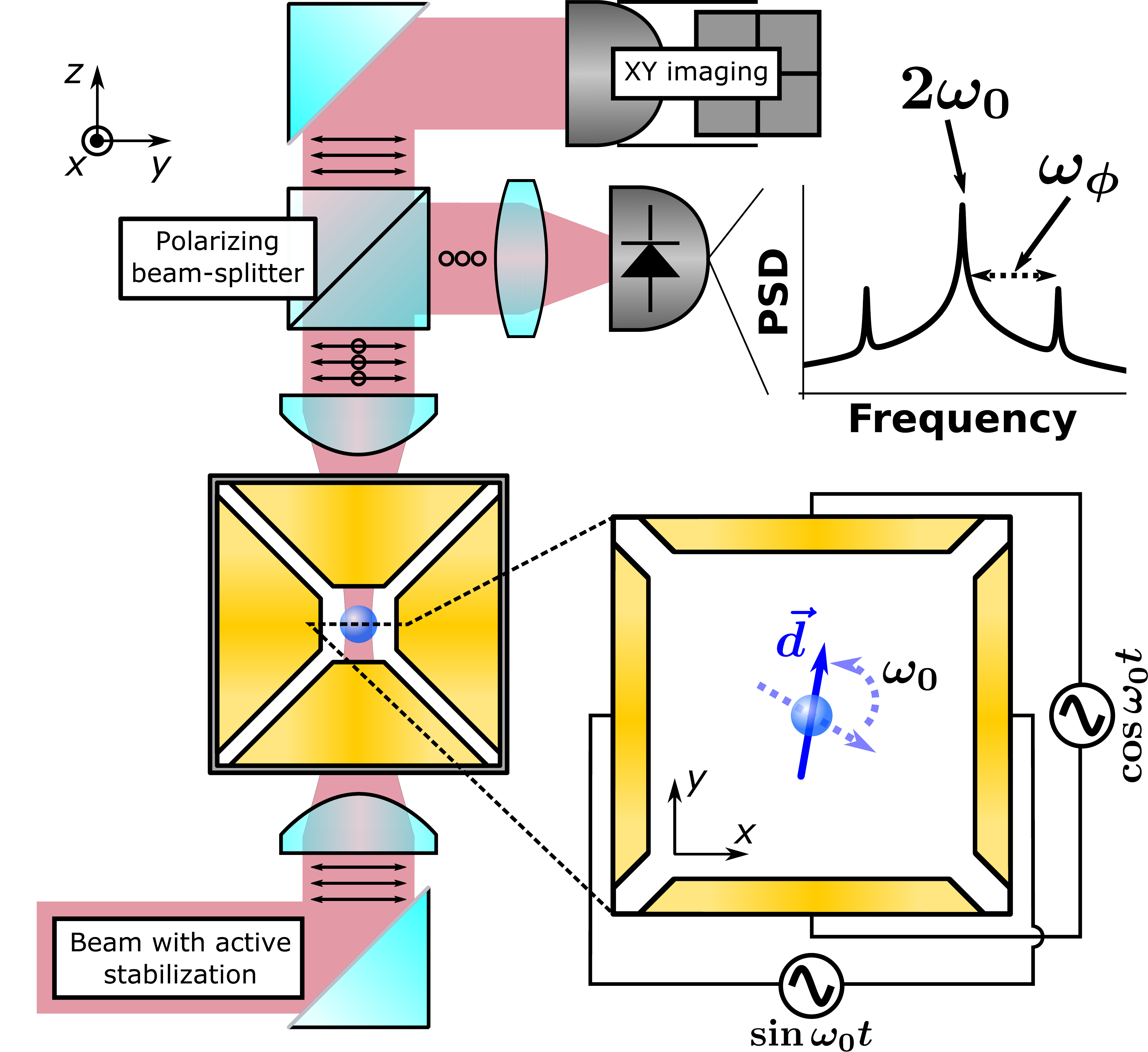}
    \caption{A schematic depiction of the central features of the apparatus including the optical trap, the surrounding electrode structure, and parts of the imaging system. The lower right inset demonstrates a typical electric field configuration when driving a trapped MS to rotate. The upper right inset is an idealized version of the expected signal from the cross-polarized light monitor as the MS is rotating with angular velocity $\omega_0$. Note that the coordinate pair relevant for the inset are different from those in the main panel. A complete description of the apparatus can be found in Ref.~\cite{Blakemore:2021b}}
    \label{fig:apparatus}
\end{figure}

Six identical electrodes form a cubical cavity around the focus of the trapping beam, whilst also allowing mechanical and optical access via central bore holes through each electrode. This architecture allows one to set both the value of the electric field and its gradient at the location of trap. By applying an individually phased sinuosidal voltage to four of the six electrodes, such that the chosen four lie within the same plane, a rotating electric field of constant magnitude can be generated. The orientation of the MS electric dipole moment then aligns with the electric field, and the angular position can be actively driven by phase-modulating the driving voltages. A schematic depiction of the optical trap is shown in Fig.~\ref{fig:apparatus}. More details on the apparatus are given in Refs.~\cite{Rider:2018, Blakemore:2021b}.

\begin{figure}[b]
    \centering
    \includegraphics[width=1\columnwidth]{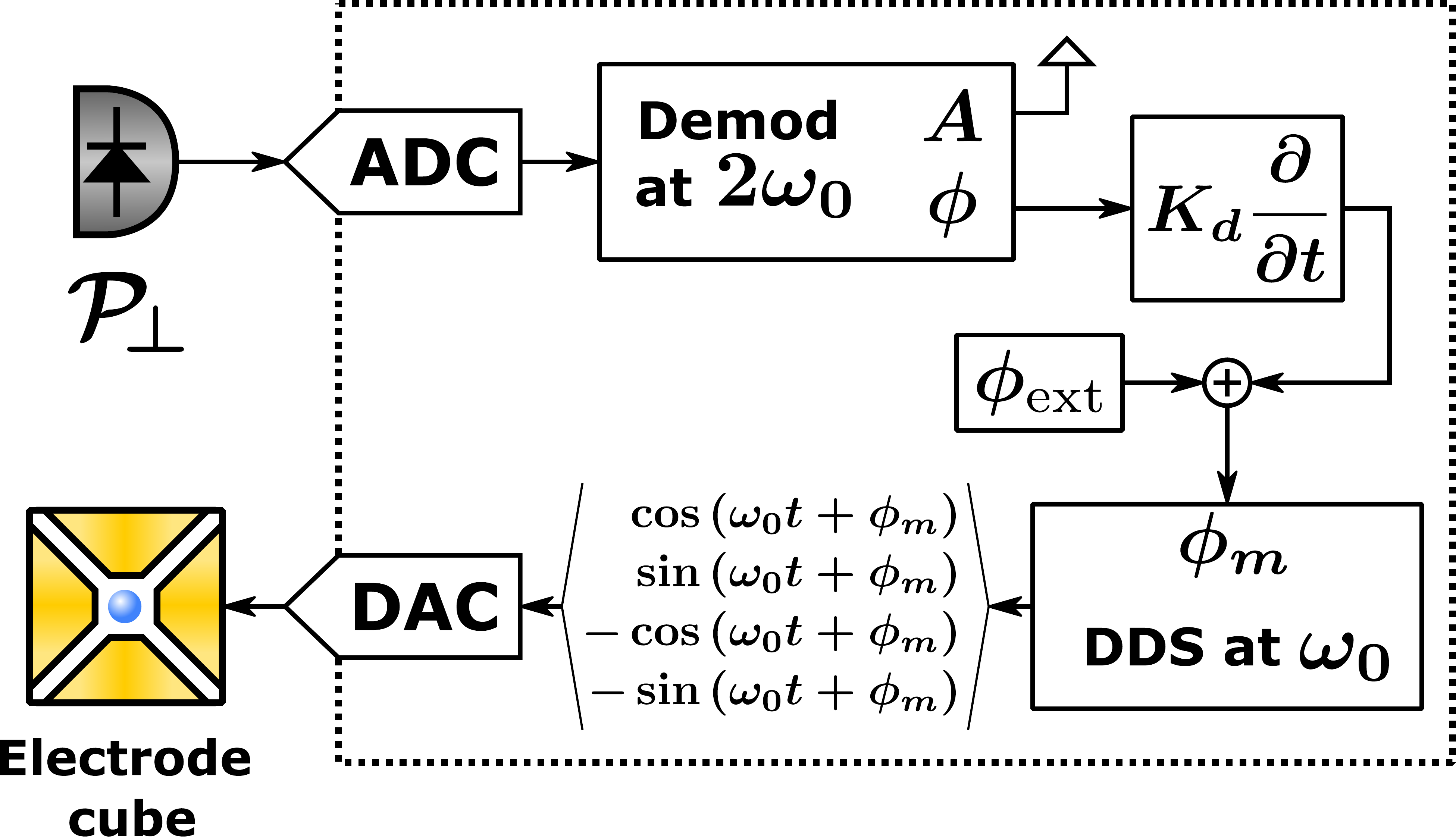}
    \caption{Block diagram of the feedback architecture, where all of the elements within the dotted border are integrated with the FPGA, such that the entire module is mutually clocked by the same top-level oscillator. The power of the cross-polarized light incident on the photodiode is modulated at angular frequency $2\omega_0$ for a MS rotating at $\omega_0$. The $2\omega_0$ carrier is demodulated by phase-locked sampling, yielding $\phi$, the angular position of the dipole moment relative to the electric field. The derivative can then be computed, and scaled by an arbitrary and user-controlled gain parameter $K_d$. The quantity $\phi_{\rm ext}$ represents an arbitrary user-defined phase that can be added to the phase modulation, $\phi_m$, of the electric field. 
    }
    \label{fig:feedback}
\end{figure}

The rotational degrees of freedom of a rotating MS are monitored by taking advantage of the residual birefringence present in St{\"o}ber process silica MSs~\cite{Monteiro:2018, Rider:2019, Blakemore:2020}. A birefringent MS couples some of the linearly polarized trapping light into the orthogonal linear polarization following the relation $\mathcal{P}_{\perp} = \mathcal{P}_0 \sin^2{(\mu/2)} \sin^2{(\theta)}$, where $\mathcal{P}_0$ is the incident power, $\mu \propto \Delta n$ is the phase retardation between the two axes of the birefringence, and $\theta$ is the angle between the projection of the fast axis into the rotation plane and the incident polarization~\cite{Jones:1941}. Thus, a birefringent MS driven to rotate with angular velocity $\omega_0$ will generate cross-polarized light with an intensity modulation at $2 \omega_0$.

The cross-polarized light is separated from the transmitted light with a polarizing beam-splitter (PBS) and projected onto a photodiode. The modulating photocurrent is first converted to a voltage, amplified, digitized, digitally filtered around $2 \omega_0$, and finally digitally-demodulated, following the technique described in Refs.~\cite{Kawasaki:2020, Blakemore:2021b}. The digitization, filtering, and demodulation operations are performed with a field-programmable gate array (FPGA, NI PCIe-7841) in order to derive the feedback signal, while the amplified photodiode output is also digitized in parallel by a second analog-to-digital converter (ADC, NI PXI-6259) operated at \SI{500}{\kilo\hertz}, and stored for offline analysis with monitor signals of the four driving voltages. A schematic view of the feedback architecture is shown in Fig.~\ref{fig:feedback}.

The drive voltages are first generated by the FPGA configured to operate as a direct digital synthesis (DDS) waveform generator. Full digital control allows the generation of four distinct signals from a single DDS: the $\sin$, $\cos$, $-\sin$, and $-\cos$ components. By phase modulating the top-level DDS, the angular position of the resultant electric field vector is necessarily modulated. While the internal structure of the DDS is clocked at \SI{40}{\mega\hertz}, the digital-to-analog converters (DACs) are updated at \SI{1}{\mega\hertz}, and for the data presented here, the rotation velocity was fixed to \SI{25}{\kilo\hertz}. The four DDS outputs are amplified (Tabor 9400) to $\mathcal{O}(\SI{100}{\volt})$ before driving the electrodes. A finite element analysis (FEA) of the electrode structure suggests that electric fields of order \SI{100}{\kilo\volt\per\meter} are possible with this hardware configuration.

Note that all components within the FPGA module are clocked by the same top-level oscillator, ensuring a phase-lock between signal generation and subsequent sampling and demodulation.

\section{THE LIBRATIONAL DEGREE OF FREEDOM}
\label{sec:libration}

Silica MS produced via the St{\"o}ber process posses an electric dipole moment $\vec{d}$~\cite{Rider:2016, Rider:2019, Afek:2021a, Afek:2021b} which, for otherwise identical spheres, can vary in magnitude by more than a factor of 10~\cite{Blakemore:2021b, Afek:2021a}. If the MS is subject to a rotating electric field of the form $\vec{E} = E \left[ \sin{(\omega_0 t)} \vec{\hat{x}} + \cos{(\omega_0 t)} \vec{\hat{y}} \right]$, where the choice of $\vec{\hat{z}}$ as the rotation axis is arbitrary, the orientation of the MS's electric dipole moment tends to align with the direction of the electric field due to the torque $\vec{d} \times \vec{E}$. Additionally, the MS is subject to a drag torque proportional to it's angular velocity, and a randomly fluctuating thermal torque, both a result of collisions with residual gas molecules. By defining $\phi'$ as the angle between the electric field vector $\vec{E}$ and the orientation of the dipole moment $\vec{d}$, and restricting the analysis to the plane of rotation ($xy$-plane), the angular momentum, $\vec{L}$ of the MS is governed by the following equation of motion:
\begin{equation} \label{eq:eom}
\frac{\partial \vec{L}}{\partial t} = \left[ \underbrace{E \, d \sin{(\omega_0 t - \phi')}}_{\substack{\text{driving torque}}} - \underbrace{\beta_{\rm rot} \frac{\partial \phi'}{\partial t}}_{\substack{\text{drag torque}}} + \underbrace{ \sqrt{S_{\rm th}} \eta(t)}_{\substack{\text{thermal torque}}} \right] \vec{\hat{z}}, \notag
\end{equation}
\noindent with $d=|\vec{d}|$ being the magnitude of the MS electric dipole moment, $\beta_{\rm rot}$ being the rotational drag coefficient of the MS, $S_{\rm th} = 4 k_B T \beta_{\rm rot}$ being the single-sided power spectral density of the thermal torque noise with $k_B$ the Boltzmann constant and $T$ the temperature of the residual gas, and with $\eta$ being a time-domain representation of a stochastic Wiener process, such that $\overline{\fft{\eta}} \fft{\eta} = 1$ with $\mathcal{F}$ being the Fourier transform operator. A derivation of $S_{\rm th}$ and $\beta_{\rm rot}$ is detailed with great care in both Refs.~\cite{Cavalleri:2010, Martinetz:2018}. 

Now transform the equation of motion to the frame co-rotating with the electric field by defining the angular coordinate $\phi = \phi' - \omega_0 t$ and recognizing that $\vec{L} = I \dot{\phi'}$ where $I$ is the MS moment of inertia. The result has an equilibrium solution, found by setting $\ddot{\phi} = \dot{\phi} = 0$ and momentarily ignoring the stochastic drive. Physically, the equilibrium solution is induced by the overall gas drag from the rotation and is given by $\phi_{\rm eq} \approx - \arcsin{(\beta_{\rm rot} \omega_0 / E \, d)}$, relative to $\phi_{\rm frame} = \omega_0 t$. The numerical value can be estimated by considering the residual gas pressure and species, in the present data, is dominated by ${\sim} \SI{2e-6}{\hecto\pascal}$ of H$_2$0, as well as typical values of the dipole moment, \SIrange{100}{2000}{\elementarycharge\micro\meter}, and chosen electric field conditions, $E \sim \SIrange{10}{100}{\kilo\volt\per\meter}$ and $\omega_0 = 2 \pi (\SI{25}{\kilo\hertz})$. We find $\phi_{\rm eq} \approx \SIrange{-1.3e-7}{-2.6e-5}{\radian}$, and thus, the constant term is dropped from the formalism.

The effect of active feedback via modulation of the phase of the rotating electric field would change the argument of the $\sin$ in Eq.~(\ref{eq:eom}) to $(\omega_0 t + \phi_m - \phi')$. Implementing pure derivative gain, to mimic the effect of damping, of the form $\phi_m = - K_d \dot{\phi}$ with $K_d$ a tunable constant, and linearizing the equation of motion, we arrive at the result:
\begin{equation} \label{eq:phi_eom}
\frac{\partial^2 \phi}{\partial t^2} + \gamma \frac{\partial \phi}{\partial t} + \omega_{\phi}^2 \phi = -K_d \omega_{\phi}^2 \left( \frac{\partial \phi}{\partial t} + \xi \right) + \frac{\sqrt{S_{\rm th}}}{I} \eta 
\end{equation}
\noindent with $\xi$ representing the measurement noise necessarily injected by the feedback, and where we have defined a damping coefficient $\gamma$ and natural frequency $\omega_{\phi}$:
\begin{equation} \label{eq:lib_freq}
\gamma \equiv \frac{\beta_{\rm rot}}{I} \quad \quad \omega_{\phi} \equiv \sqrt{\frac{E \, d}{I}}
\end{equation}
\noindent recognizing the usual equation of motion for a damped harmonic oscillator with forcing terms. Computational delays inherent to the feedback architecture are such that the calculated value of $\dot{\phi}$ in the feedback forcing term is in fact delayed in time, i.e. $\dot{\phi} = \dot{\phi}(t-t_{\rm fb})$ for the term proportional to $K_d$, with $t_{\rm fb} \sim \mathcal{O}(\SI{100}{\micro\second})$.

\subsection{Step Response - Homogeneous Solution}

Consider the response of the system to a step function, such as would result from a discrete change in the orientation of the rotating electric field. If the step is sufficiently fast compared to the length of one librational period, it can be modeled as an instantaneous effect. Further ignoring the thermally driven portion of the solution and assuming that any transients have been fully damped, the step response can be derived by integrating Eq.~(\ref{eq:phi_eom}) subject to the initial conditions $\phi(0) = \phi_0$ and $\dot{\phi}(0) = 0$, where a step of magnitude $\phi_0$ is assumed at time $t=0$.

A proper treatment would consider the full functional form of the potential well ($U \sim \int \sin{\phi} \, d\phi$ instead of $\sim \phi^2$), include anomalous dissipation generated by the feedback-injected noise $\xi$, and account for the causal limitation represented by $t_{\rm fb}$, but for the purpose of this first demonstration of librational cooling, we appeal to the approximated case with a quadratic potential and with $\xi=0$ as well as $t_{\rm fb} = 0$. The accuracy and limitations of the approximation will be discussed below. Integrating the simplified equation of motion $\ddot{\phi} + (\gamma + k_d) \dot{\phi} + \omega_{\phi}^2 \phi = 0$, subject to the aforementioned initial conditions and with $k_d \equiv K_d \omega_{\phi}^2$ defined for brevity, $\phi(t)$ is obtained as:
\begin{equation} \label{eq:transient_soln}
\phi(t) = \phi_0 e^{-\gamma_d t / 2} \cos{\left( \sqrt{\omega^2 - \frac{\gamma_d^2}{4}} \, t \right)}
\end{equation}
\noindent where $\gamma_d \equiv \gamma + k_d$. The time-constant of the exponential envelope $\tau = 2 / \gamma_d$ is determined by the combined effect of the system's intrinsic damping $\gamma$, assumed to be dominated by collisions with residual gas, and the feedback-induced damping $k_d$, allowing the effect of the feedback to be quantified.

The analytic solution given by Eq.~(\ref{eq:transient_soln}) can be evaluated against a numerical solution obtained by integrating the full form of the sinusoidal potential. A Runge-Kutta integrator was used for this purpose, with a few different electric field amplitudes and effective damping coefficients $\gamma_d$, chosen to span the approximate range of both parameters. A comparison between the exponential fits to the decaying envelopes of oscillation for both the analytic solution of the approximate potential, and the numerical solution of the full potential (with otherwise identical $E$, $d$, $I$, $\gamma_d$) is used to estimate the systematic bias associated with the approximated solution above. It is found that a fixed multiplicative factor can account for this bias: $(\hat{\tau})_{\rm truth} \approx (\hat{\tau})_{\rm linear} \approx 1.1 (\hat{\tau})_{\rm full}$. Thus, a linear scaling of the observed $\gamma_d$ with increasing derivative gain, derived from a na{\"i}ve exponential fit to actual data, is expected in the analysis below.

\subsection{Thermal Steady State}

The thermally-driven steady state motion of the MS librational degree of freedom should also depend on the level of the applied feedback. An expression for the expected power spectral density of the librational motion can be derived by considering the Fourier transform of Eq.~(\ref{eq:phi_eom}), with $k_d = K_d \omega_{\phi}^2$ as before. We generally follow the extensive formalism presented in the Supplementary Material of Ref.~\cite{Conangla:2019}. Let $\tilde{\phi}(\omega) \equiv \fft{\phi(t)}$ and recall that for individual Fourier component solutions, we know $\fft{\phi(t - t_{\rm fb})} = e^{- i \omega t_{\rm fb}} \tilde{\phi}$ and $\fft{\dot{\phi}} = i \omega \tilde{\phi}$, with $t_{\rm fb}$ as before. We find that:
\begin{equation} \label{eq:phi_transform}
\tilde{\phi} = \frac{\sqrt{S_{\rm th}}/I - i k_d \omega \tilde{\xi}}{\left[ \omega_{\phi}^2 - \omega^2 \right] + i \omega \left[ \gamma + k_d e^{- i \omega t_{\rm fb}}  \right]},
\end{equation}
\noindent with $\tilde{\xi}$ being the Fourier transform of the measurement noise. We can compute the expected power spectral density of the librational motion directly: 
\begin{equation} \label{eq:phi_psd}
\begin{split}
S_{\phi\phi} &= \overline{ \tilde{\phi} } \tilde{\phi}  \\
&= \frac{S_{\rm th}/I^2}{G(\omega)} + \frac{k_d^2 \omega^2 S_{\xi\xi} }{G(\omega)},
\end{split}
\end{equation}
\noindent with denominator $G(\omega) = [ \omega_{\phi}^2 - \omega^2 + \omega k_d \sin{(\omega t_{\rm fb})} ]^2 + \omega^2 [ \gamma + k_d \cos{(\omega t_{\rm fb})} ]^2$, and where it has been assumed that $\xi$ and $\eta$ are uncorrelated.

Solving for the quantity $\tilde{\phi} + \tilde{\xi}$, we can compute the expected power spectral density of the libration, as observed from the in-loop detector:
\begin{equation} \label{eq:phi_psd_il}
\begin{split}
S_{\rm IL} &= \overline{( \tilde{\phi} + \tilde{\xi} )} ( \tilde{\phi} + \tilde{\xi}) \\
&= \frac{S_{\rm th}/I^2}{G(\omega)} + \frac{[ (\omega_{\phi}^2 - \omega^2)^2 + \gamma^2 \omega_{\phi}^2 ] S_{\xi\xi} }{G(\omega)},
\end{split}
\end{equation}
\noindent with $G(\omega)$ as before.

The intended effect of the feedback is to introduce additional damping. However, due to the fixed temporal phase shift $t_{\rm fb}$ associated to this particular feedback implementation, the effect on the observed power spectral density is nontrivial. Importantly, there is a noise injection term proportional to $S_{\xi\xi}$, but suppressed at the resonance, and the denominator $G(\omega)$ induces a clear asymmetry in the observed spectral density. The former effect is the well-understood result of noise cancellation in the detector induced by the feedback system, and is often referred to as noise squashing~\cite{Tebbenjohanns:2019, Conangla:2019, Monteiro:2020}. Systems with both in-loop and out-of-loop detectors can circumvent the noise squashing, but the signal level in this first iteration of librational cooling was not sufficient for distribution to multiple detectors. 

The parameter $k_d$ can be extracted by fitting Eq.~(\ref{eq:phi_psd_il}) to the observed spectra, and together with the known values of $I$, $\omega_{\phi}$, and $\gamma$, an effective temperature of the librational motion can be estimated. From equipartition and Parseval's theorem, maintaining the convention of single-sided PSDs, the effective mode temperature can be calculated:
\begin{equation} \label{eq:t_eff}
\begin{split}
k_B T_{\rm eff} &= I \omega_{\phi}^2 \langle \phi^2 \rangle \\
&= I \omega_{\phi}^2 \frac{1}{2\pi} \int_{0}^{\infty} S_{\phi\phi} d\omega \\
&\approx \frac{I \omega_{\phi}^2}{4} \left[ \frac{S_{\rm th}/I^2}{ \omega_{\phi}^2 (\gamma + k_d)} + \frac{k_d^2 S_{\xi\xi}}{\gamma + k_d} \right]
\end{split}
\end{equation}
\noindent where in the final line it has been assumed that $t_{\rm fb} = 0$ in order arrive at a closed form expression for $T_{\rm eff}$.  Clearly, any estimation via the expression in Eq.~(\ref{eq:t_eff}) will have limited accuracy since it is known that $t_{\rm fb} \neq 0$. The acquired spectra will be numerically integrated in future iterations of the apparatus, taking advantage of both improved signal-to-noise and a dedicated out-of-loop sensor.

\section{RESULTS}
\label{sec:results}

Librational feedback cooling was demonstrated with three distinct MSs, all from the same lot with diameter \SI{4.70\pm0.08}{\micro\meter} and each trapped for approximately one month. A variety of different derivative gain values, $K_d$, were used, as well as few different electric field amplitudes so that effective values of $k_d$ span roughly 4 orders of magnitude. The degree of cooling for a specific choice of parameters was quantified via two distinct methods: application of a step, and thermalization.

\subsection{Discrete Phase Step}
\label{sec:step}

As the driving voltages that source the rotating electric field are generated by a single top-level DDS, it is possible to apply arbitrary phase steps by propagating the phase of each of the four output sinusoids simultaneously. As a result, the applied electric field rapidly changes orientation. Electric field phase changes of $\Delta \phi = \pm \pi / 2$ in this system, corresponding to drive voltage amplitude changes of $(1/2)$ the peak-to-peak voltage, have been measured to have rise times $t_{\rm rise} < \SI{2}{\micro\second}$. This is consistent with both the \SI{50}{\kilo\ohm} termination resistance and \SI{30}{\pico\farad} electrode-to-ground capacitance as well as the ${\sim}\SI{500}{\kilo\hertz}$ full-scale bandwidth of the driving amplifier. The frequency of libration can be controlled by tuning the electric field, and usually has values $\omega_{\phi} = \sqrt{E \, d / I} \sim \mathcal{O}(2 \pi \times \SI{1}{\kilo \hertz})$, so that $\omega_{\phi} t_{\rm rise} \lesssim \SI{0.01}{\radian}$, and thus the finite rise time of the step has a negligible effect on the dynamics. This is sufficiently fast that it is effectively instantaneous relative to the $\mathcal{O}(\SI{1}{\kilo\hertz})$ fundamental frequency of the libration.

\begin{figure}[t]
    \centering
    \includegraphics[width=1\columnwidth]{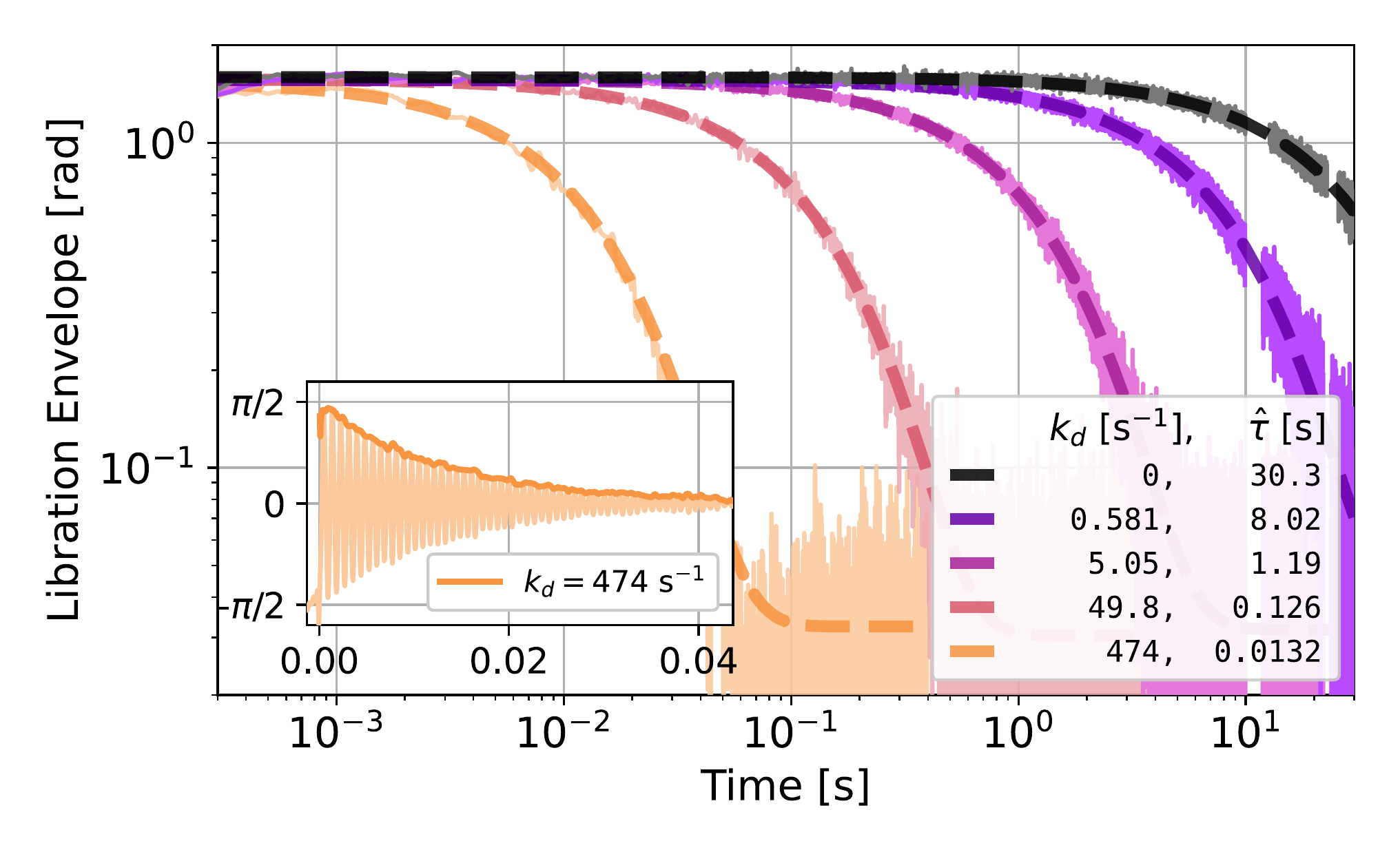}
    \caption{The amplitude envelope of the libration in response to an applied step, for a variety of derivative gain values, including $k_d = 0$ (color online). A dashed line indicates the result of exponential fitting, with the extracted damping time showing in the legend. (inset) an example of the underlying oscillation of the librational motion for the largest value of derivative gain. Some filtering artifacts are present immediately following the step, and are excluded from the exponential fit.}
    \label{fig:ringdowns}
\end{figure}

For the measurements presented here, a step of $\Delta \phi = \pm \pi / 2$ was applied to the rotating electric field (where the $\pm$ indicates that the phase offset was applied alternately in the `forward' and `backward' directions), and the subsequent ringdown of the MS's librational motion was observed. Between successive measurements, the feedback gain and electric field amplitude were first altered to their new values, and then the motion was allowed to thermalize for $> \SI{3000}{\second}$, following the expected torsional damping times (dependent on the base pressure of the vacuum chamber) observed previously on the same system~\cite{Blakemore:2020}. A few examples of the measured response to such a step are shown in Fig.~\ref{fig:ringdowns}, for one specific MS. 

The libration is extracted from the cross-polarized light signal, first by Hilbert transforming the primary rotation signal at $2 \omega_0 = 2 \times 2 \pi (\SI{50}{\kilo\hertz})$, which yields $2\phi'$, the angular position of the MS in the \emph{lab frame}. Using the known drive frequency $\omega_0$, the libration is then reconstructed as $\phi = \phi' - \omega_0 t$. Finally, the amplitude envelope of the libration is extracted by a second Hilbert transform of the reconstructed $\phi$.

\begin{figure}[t]
    \centering
    \includegraphics[width=1\columnwidth]{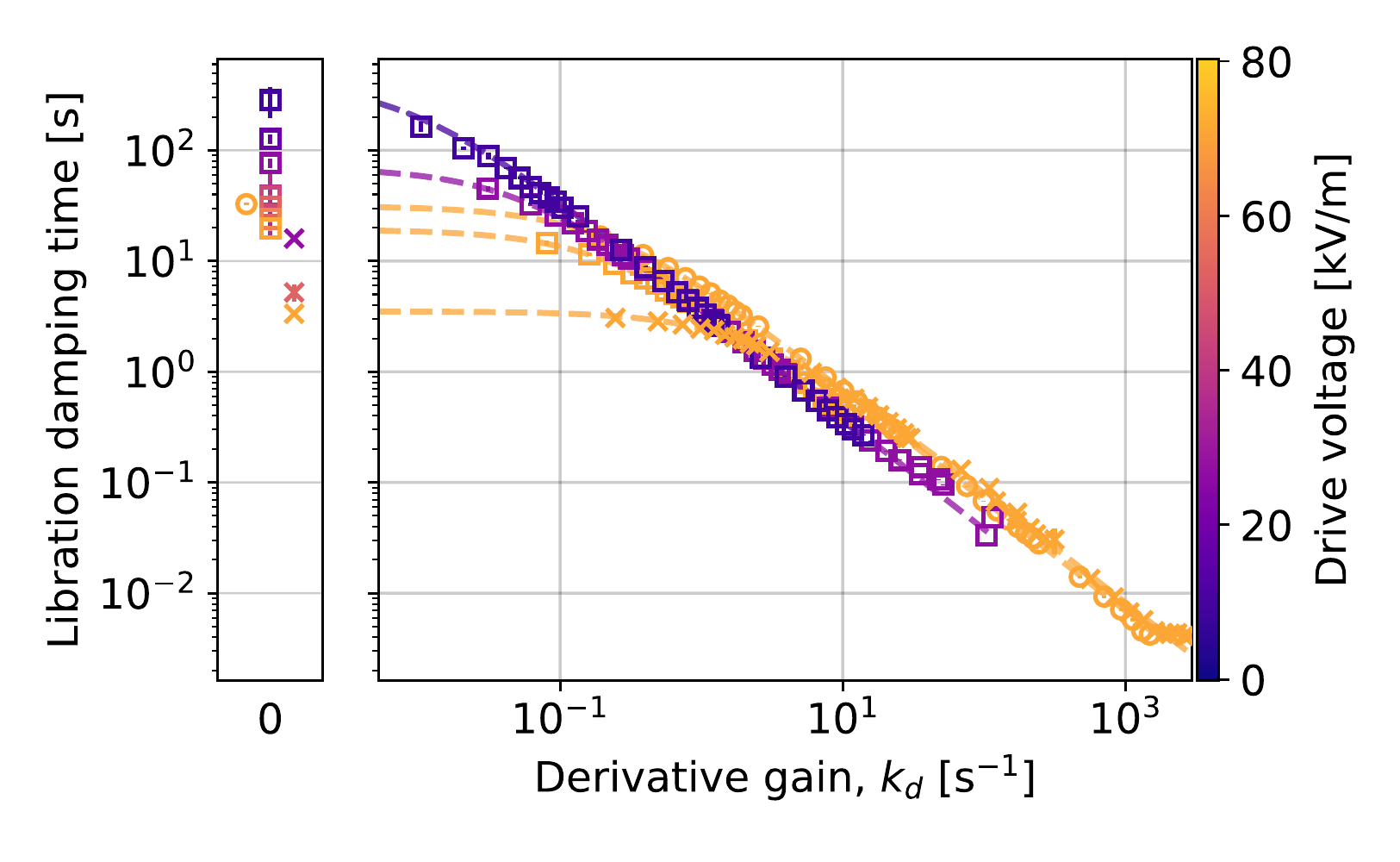}
    \caption{Summary of the libration phase step measurements. The damping time extracted from exponential fits is plotted as a function of the applied derivative gain. Different colors indicate different electric field amplitudes, while different marker shapes indicate distinct MSs. (left) Damping times in the absence of applied feedback, showing a clear dependence on both drive amplitude and MS, with the former being a monotonic relation. Data from distinct MSs has been offset horizontally to aid their visibility. (right) damping times with feedback on. Dashed lines indicate fits to the expected scaling relation $\hat{\tau} = 2 / (\gamma + C k_d)$, where $C$ is an arbitrary scaling constant found to be necessary to match the observed relation between $k_d$ and $\hat{\tau}$.}
    \label{fig:ringdown_summary}
\end{figure}

A damping time is inferred by downsampling the amplitude envelope, and then fitting the result with a decaying exponential, including a constant offset to account for the RMS amplitude of the thermal motion after the transient response has fully decayed. The fit of the amplitude envelope is constrained to the domain $[t_0 + 0.1\tau_0, t_0 + 2.0\tau_0]$ where $t_0$ indicates when the step was applied and $\tau_0$ is an initial estimate of the $e$-folding time obtained from the mean of data samples that cross $\phi \sim (\pi / 2)e^{-1}$ after the step. From Sec.~\ref{sec:libration} and the aforementioned numerical integrations of the full potential (rather than the harmonic approximation), we know that the decay time, $\tau \approx 2 / \gamma$, will be systematically underestimated by the na{\"i}ve exponential fit to actual data, but by a fixed multiplicative constant. Regardless, the scaling of the decay time from the fit $\hat{\tau}$ as a function of $k_d$ still allows characterization of the cooling.

A summary of all step response measurements is shown in Fig.~\ref{fig:ringdown_summary}, with the extracted $\hat{\tau}$ plotted as a function of the derivative gain. When the derivative gain is sufficiently small, there is some intrinsic damping that dominates. Interestingly, the zero-feedback damping times observed, $\mathcal{O}(\SIrange{10}{100}{\second})$, are inconsistent with the expected value of $\tau \approx 2 / \gamma \approx 2 I / \beta_{\rm rot} \sim \SI{4000}{\second}$~\cite{Rider:2019, Blakemore:2020} given the base pressure achieved in this vacuum system. This may be the result of phase noise either in the top-level DDS sourcing the electric field, or in the synchronization of the various ADCs, DACs, and digital demodulation operations. It may also be symptomatic of non-linearities, such as those that might arise from higher-order charge distributions present in the MS.

From Sec.~\ref{sec:libration}, the exponential damping time might be expected to follow the relation $\hat{\tau} = 2 / (\gamma + k_d)$. The data are inconsistent with this expectation, instead being described by the relation $\hat{\tau} \sim 2 / (\gamma + C k_d)$ with $C$ a positive scaling constant. The constant $C$ is different for distinct MSs, but consistent across drive voltage for each MS. For the three whose data are presented here: $C_1 = 0.29 \pm 0.03 $, $C_2 = 0.53 \pm 0.05 $, and $C_3 = 0.24 \pm 0.03$. As was noted in Sec.~\ref{sec:libration}, a systematic bias in the estimate of $\hat{\tau}$ is expected from fitting a na{\"i}ve exponential to the ringdowns within the sinusoidal potential, but this factor of $\hat{\tau} \rightarrow 1.1 \hat{\tau}$ does not explain the observed deviation, nor would it be different between MSs, under the construction presented here. This is discussed further in the next section.

\subsection{Thermalization}
\label{sec:thermalization}

Immediately prior to a phase step measurement, but after the long \SI{3000}{\second} thermalization time, \SI{200}{\second} of librational motion are monitored and digitized, in 20 continuous \SI{10}{\second} integrations. For each integration, the libration $\phi$ is extracted with the same Hilbert transform discussed in the previous section, and the power spectral density is estimated by squaring the Fourier transform of the digitized signal. The observed motion is sufficiently small so that the approximation $\sin{\phi} \approx \phi$ is valid.

\begin{figure}[t]
    \centering
    \includegraphics[width=1\columnwidth]{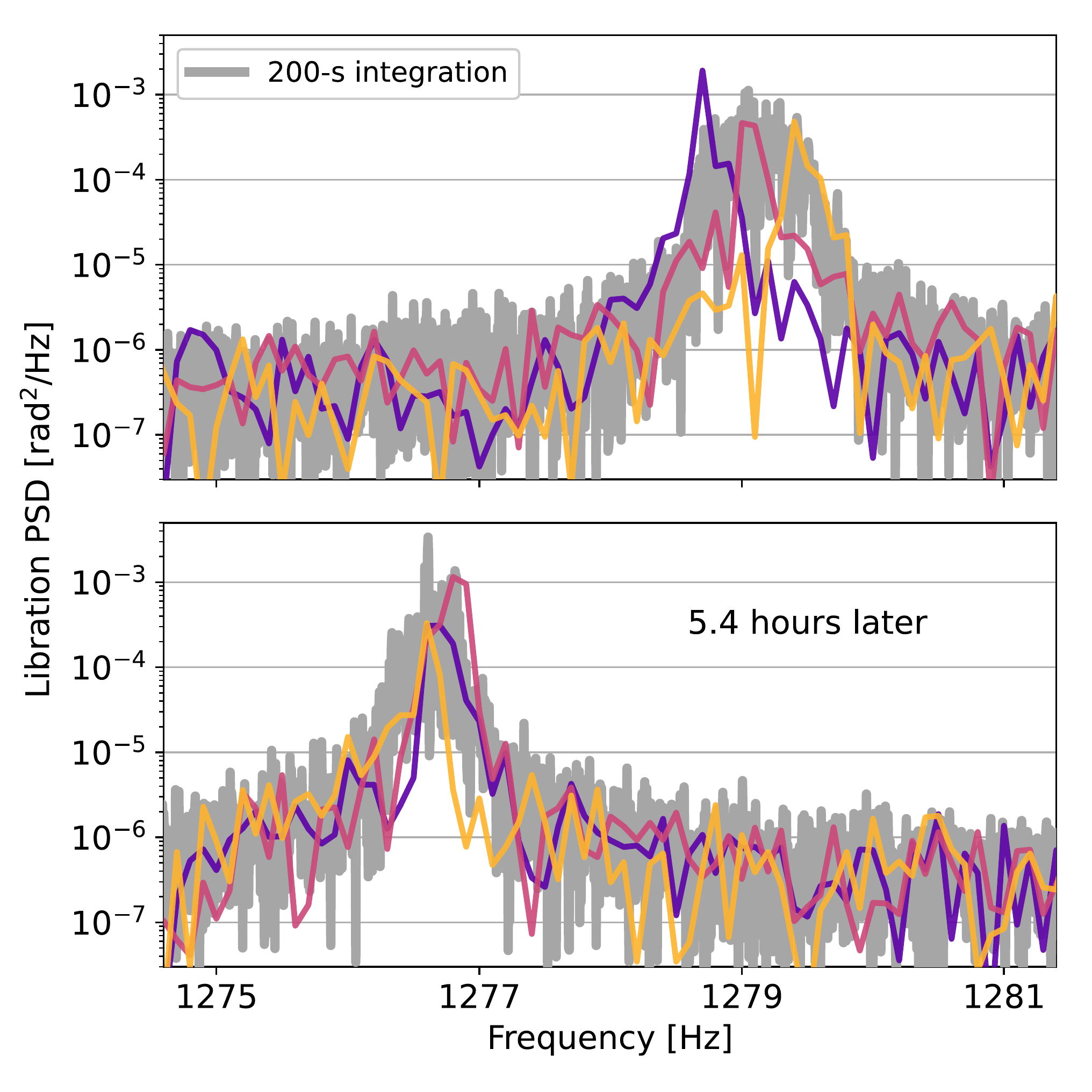}
    \caption{PSDs of the librational motion depicting a drifting central frequency. Each panel contains one \SI{200}{\second} dataset, with \SI{5.4}{\hour} between the two. The power spectral density of each full dataset is shown in gray, while the power spectral densities of individual \SI{10}{\second} integrations within the datasets are shown in curves of varying brightness and color (color online). It appears that some of the observed spectral width is driven by slow fluctuations in the center frequency, distinct from additive dissipation and subsequent line-broadening.}
    \label{fig:spectra_drift}
\end{figure}

An exemplary pair of two such \SI{200}{\second} datasets with one MS are shown in Fig.~\ref{fig:spectra_drift}, where the PSD of the individual \SI{10}{\second} are shown, as well as the PSD of the concatenated signal. Between successive integrations, the central frequency of the librational motion $\omega_{\phi}$ appears to change by $\mathcal{O}(\SI{0.1}{\hertz})$, and as a result, the PSD of the concatenated signal appears anomalously broadened. Furthermore, under identical conditions, but ${\sim} \SI{5}{\hour}$ later, the central frequency continued to drift by $>\SI{2}{\hertz}$. Both effects can hinder the estimations of damping, given that $\hat{\gamma} = \gamma - k_d$ is extracted from the width of the observed spectral feature. Dividing the data into smaller blocks, effectively implementing shorter intergration times, provides little improvement given the fixed sampling rate and frequency resolution implied the Nyquist-Shannon sampling limit~\cite{Nyquist:1928,Shannon:1949}.

In order to mitigate the effect of this drift, first, the central frequency of libration $\hat{\omega}_i$, of each of the $i$ integrations is estimated by fitting the spectral feature to Eq.~(\ref{eq:phi_psd}) with $S_{\xi} = 0$ and $t_{\rm fb} = 0$, but an added constant offset to account for detector noise. From these estimations, a mean central frequency can be defined $\overline{\omega} \equiv (1/N)\sum_i^N \hat{\omega}_i$. The librational motion in each integration is then frequency shifted by $\Delta \omega = \overline{\omega} - \hat{\omega}_i$. The frequency shift is accomplished by assuming the filtered signal $\phi(t)$ can be represented by $\phi(t) = A(t) \cos{[\omega_i t + \theta(t)]}$, as well as the implicit requirement that $A(t)$ and $\theta(t)$ change slowly relative to $\omega_i$. The analytic representation of the signal can then be constructed from the Hilbert transform, $\mathcal{H}$, as $\phi_a(t) = \phi + i \mathcal{H}[\phi] = A(t) \exp{\{ i [\omega_i t + \theta(t)]\}}$, so that a frequency shift can be implemented simply via the multiplication $\phi_a \cdot \exp{(i \Delta \omega t)}$. 

\begin{figure}[t]
    \centering
    \includegraphics[width=1\columnwidth]{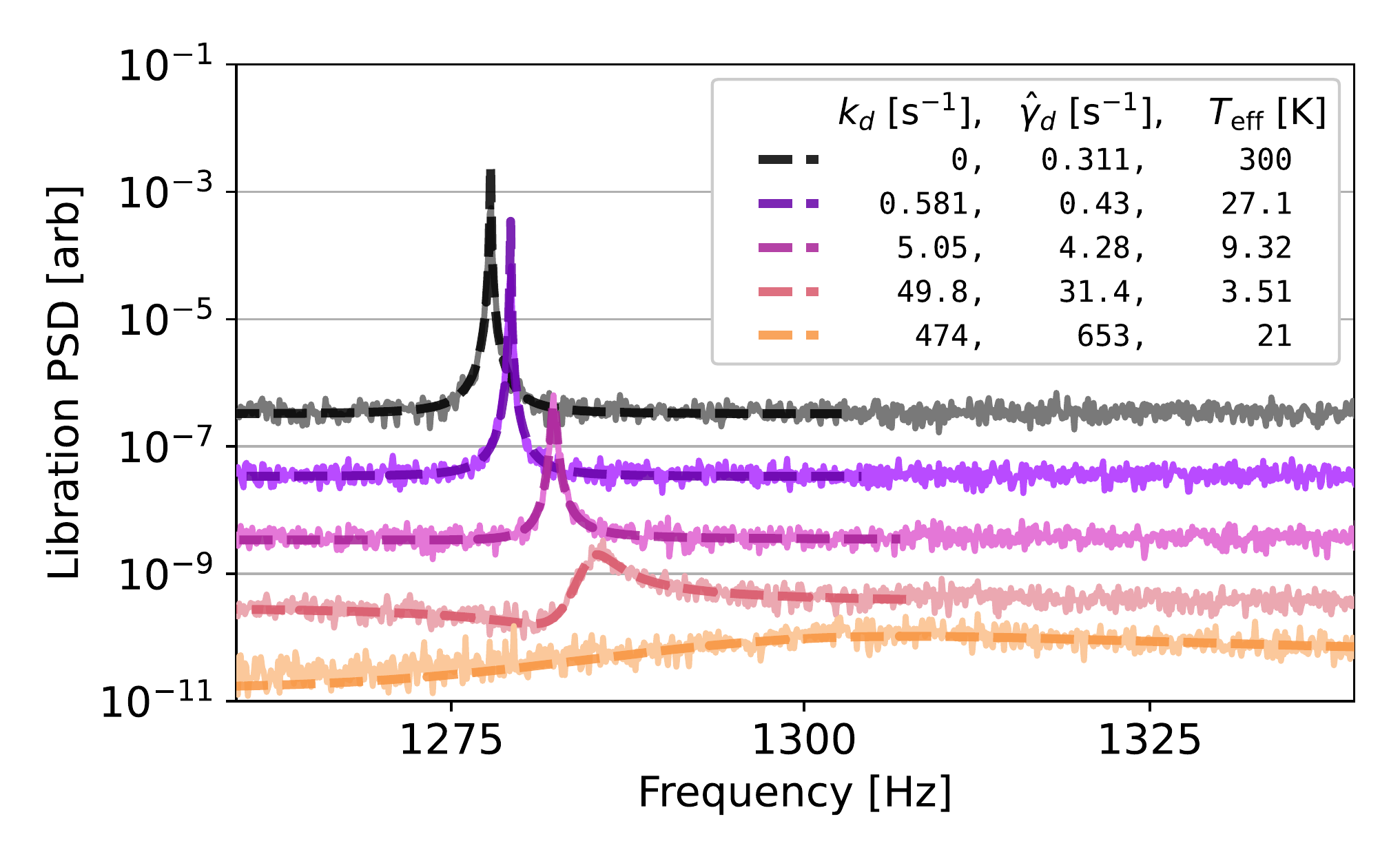}
    \caption{A few examples of the mean PSDs of the librational motion of one particular MS, for a few different values of derivative gain. Each PSD shown is the mean of the PSDs of 20 individual \SI{10}{\second} integrations, following the averaging procedure discussed in the text. The dashed lines indicates fits of the PSD to the expression in Eq.~(\ref{eq:phi_psd_il}), with the extracted value of $\hat{\gamma}_d = \gamma + k_d$ and the effective mode temperature from Eq.~(\ref{eq:t_eff}) shown in the legend.}
    \label{fig:spectra_examples}
\end{figure}

\begin{figure}[t]
    \centering
    \includegraphics[width=1\columnwidth]{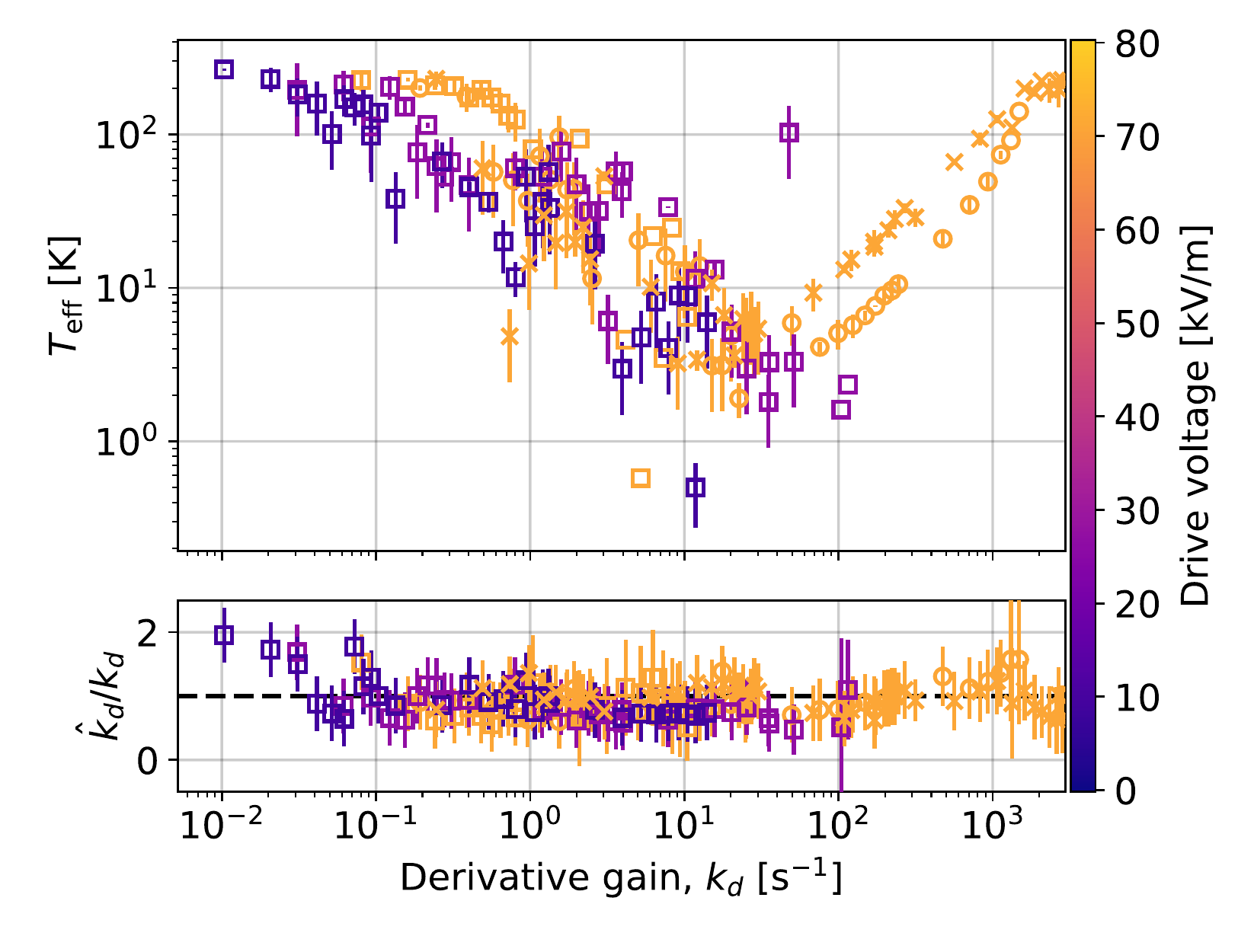}
    \caption{Summary of the libration thermalization measurements, where, as before, different colors indicate different electric field amplitudes, while different marker shapes indicate distinct MSs. (upper) Effective mode temperature calculated via Eq.~(\ref{eq:t_eff}), with the values of $\omega_{\phi}$, $\gamma$, $\hat{k}_d$, $S_{\rm th}$, and $S_{\xi}$ extracted from fitting Eq.~(\ref{eq:phi_psd_il}) to the observed PSDs. Relatively large error bars are indicative of the high degree of correlation between various fitting parameters, as well as the limited signal to noise ratio. The re-heating observed for large $k_d$ is consistent with broadband noise injection from the feedback loop as the gain is increased. (lower) Ratio of the extracted value of $\hat{k}_d$, relative to the expected value $k_d = K_d \omega_{\phi}^2$.}
    \label{fig:thermalization_summary}
\end{figure}

The real part of the frequency-shifted analytic signal is then the desired librational motion, which is constructed separately for each individual integration. The average PSD of all such integrations for one \SI{200}{\second} measurement is fit to Eq.~(\ref{eq:phi_psd_il}) in order to estimate $k_d$ and $S_{\xi}$, where the value of $\gamma$ is loosely constrained to $2 / \tau(k_d = 0)$, i.e. the level of instrinsic damping observed during ringdown measurements when the effect of the active feedback is negligible. A few examples of the average PSDs together with their fits are shown in Fig.~\ref{fig:spectra_examples}, and a summary of all the fitting results in Fig.~\ref{fig:thermalization_summary}, with $T_{\rm eff}$, calculated from Eq.~(\ref{eq:t_eff}), plotted as a function of $k_d$.

\subsection{Anomalous dissipation}

The formalism presented in Sec.~\ref{sec:libration} suggests that $\omega_{\phi} = \sqrt{E \, d / I}$, which would normally be assumed constant for a fixed electric field magnitude. Clearly, the measurements presented here are inconsistent with that assumption, and there is not only a source of anomalous dissipation for the librational degree of freedom, as was shown in Sec.~\ref{sec:step}, but also slow drifts in the central frequency, as seen in Sec.~\ref{sec:thermalization}. The magnitude of the electric field is measured to be constant within $\pm \SI{50}{\volt\per\meter}$, consistent with the absolute accuracy of the DACs sourcing it, ${\sim}0.03\%$ of their full scale, as reported by the manufacturer over both \SI{24}{\hour} and 1~year timescales~\cite{ni_pcie_7841}.

This naturally implies that some combination of the MS electric dipole moment $d$ and moment of inertia $I$ are fluctuating. For the MSs used~\cite{bangs_laboratories}, electric dipole moments over the range \SIrange{100}{2000}{\elementarycharge\micro\meter} have been observed~\cite{Blakemore:2021b}. The underlying mechanism that gives rise to these dipole moments is not fully understood and it has been observed that ionization of residual gas in close proximity to a trapped MS can greatly affect the magnitude of the dipole moment~\cite{Blakemore:2020}. 

It is natural then to suggest that both the the anomalous dissipation and slow drifts of the central frequency are a result of a changing charge multipole within the MS. Using one of the three MSs presented in this work, two dedicated dipole moment measurements following the procedure first established in Ref.~\cite{Rider:2019} and separated by $\mathcal{O}(1~\mathrm{month})$ yielded \SI[parse-numbers=false]{1804 \pm 39 (stat.) \pm 84 (sys.)}{\elementarycharge\micro\meter} initially, and then \SI[parse-numbers=false]{1094 \pm 24 (stat.) \pm 51 (sys.)}{\elementarycharge\micro\meter}, a very significant change for what has often been assumed a persistent physical characteristic of the MSs. Indeed, multiple physics searches with this and similar apparatuses have encountered systematic effects consistent with electromagnetic interactions that slowly fluctuate in time~\cite{Blakemore:2021a, Priel:2022}.

\section{CONCLUSION}
\label{sec:conclusion}

We have successfully demonstrated feedback cooling of a librational degree of freedom of an optically trapped silica microsphere in vacuum. In this first implementation of librational cooling, feedback was accomplished primarily with damping constructed from the derivative of the libration. The level of feedback was tuned over roughly four orders of magnitude for an individual microsphere, characterized by applying both a step and observing the resulting transient, as well as analyzing the steady-state motion once it has thermalized with the environment. Transient damping times scale inversely with the applied derivative gain, as expected, although with an overall systematic bias that is distinct across individual microspheres, and is as of yet unexplained. Thermally-driven power spectral densities of the libration have widths consistent with the applied derivative gain when the latter is sufficiently large.

Both the step response and steady-state measurements presented suggest some source of anomalous dissipation in this system, as well as drifts of the physical properties of optically trapped microspheres. The librational damping time and thermal spectral width in the absence of feedback are inconsistent with expectation from fluctuations driven by residual gas, although this is likely a symptom of noise in the driving the electronics given the observed scaling with drive amplitude, and furthermore it can be tested in the future with improved hardware. At the same time, large drifts in the fundamental frequency of the libration are observed, suggesting the ratio $(d/I)$ is changing by up to a factor of two. Any such changes would have consequences for precision measurements limited by electrostatic backgrounds.

\begin{acknowledgments}

This work was supported, in part, by ONR grant N00014-18-1-2409, NSF grant PHY1802952, and the Heising-Simons Foundation. C.P.B. acknowledges the partial support of a Gerald~J. Lieberman Fellowship of Stanford University. A.K. acknowledges the partial support of a William~M. and Jane~D. Fairbank Postdoctoral Fellowship of Stanford University. N.P. acknowledges the partial support of the Koret Foundation. We acknowledge regular discussions on the physics of trapped microspheres with the group of Prof. D.~Moore at Yale.  

\end{acknowledgments}

\bibliography{librational_cooling}

\end{document}